\input{aipcheck}
\documentclass[,final]{aipproc}

\layoutstyle{8x11single}

\newcommand{\cbar}{{\bar{c}}}
\newcommand{\ccbar}{{c\cbar}}

\newcommand{\jpsi}{{\rm{J}/\psi}}

\newcommand{\pp}{{\rm{p}\rm{p}}}
\newcommand{\pbpb}{{\rm{Pb-Pb}}}

\newcommand{\auau}{{\rm{Au-Au}}}

\newcommand{\ncol}{{N_{\rm{coll}}}}
\newcommand{\npart}{{N_{\rm{part}}}}
\newcommand{\rcp}{{R_{\rm{CP}}}}
\newcommand{\raa}{{R_{\rm{AA}}}}

\newcommand{\pt}{{p_{\rm T}}}

\newcommand{\snn}{{\sqrt{s_{\rm NN}}}}

\begin{document}

\title{$\jpsi$ production in $\pp$ and $\pbpb$ with ALICE at the LHC}


\author{H. Pereira Da Costa, for the ALICE Collaboration}{
  address={CEA/IRFU/SPHN, L'Orme Des Merisiers, 91191 Gif/Yvette CEDEX, FRANCE}
}

\begin{abstract}
ALICE is the Large Hadron Collider (LHC) experiment dedicated to the study of heavy ion collisions. The main purpose of ALICE is to investigate the properties of a new state of deconfined nuclear matter, the Quark Gluon Plasma (QGP). Quarkonium measurements will play a crucial role in this investigation due to the interplay of several competing mechanisms that are predicted to modify its production in the presence of a QGP. During the 2010 and 2011 LHC campaigns, ALICE took $\pp$ data at $\sqrt{s}=2.76$ and $7$~TeV and $\pbpb$ data at $\snn=2.76$~TeV. We present the latest results of $\jpsi$ production under these conditions, measured by the ALICE experiment at both mid- and forward-rapidities.
\end{abstract}
\maketitle

The $\jpsi$ meson has long been considered a favored probe to study the formation of a Quark Gluon Plasma (QGP) in heavy ion collisions and has therefore been extensively studied experimentally, notably at the SPS (CERN) and at the RHIC (BNL). It is also studied at the LHC at unprecedentedly high collision energies. The $\jpsi$ production was originally predicted to be suppressed in presence of a QGP with respect to its production in $\pp$ collisions via a color screening mechanism similar to the Debye screening in QED~\cite{Satz:1986}, provided that the temperature of the QGP is high enough. At the LHC energies however, the $\jpsi$ production could also be enhanced due to the coalescence of uncorrelated $\ccbar$ pairs from the hot medium\cite{Andronic:2003,Svetitsky:1988,Thews:2005}. Additionally, several mechanisms can also modify the production of $\jpsi$ mesons in heavy ion collisions even in absence of a QGP, such as the modification of the parton distribution functions of the nucleon inside a nucleus. Measuring the $\jpsi$ production in $\pp$ collisions is mandatory as it serves as a reference for studying its modification in heavy ion collisions. It is also interesting on its own since the $\jpsi$ production mechanism, more precisely the mechanism by which the initial $\ccbar$ pair produced by the hard scattering of two partons is neutralized to form a $\jpsi$ meson, is still largely not understood. Precise measurements performed at the LHC energies will provide additional constrains to the models that aim to describe this mechanism.

The ALICE experiment~\cite{ALICE:2008} measures $\jpsi$ mesons down to zero transverse momentum at both mid- ($|y|<0.9$) and forward- ($2.5<y<4$) rapidities. At mid-rapidity, $\jpsi$ mesons are measured through their decay into two electrons. Two detectors are used for this analysis: the Time Projection Chamber (TPC) and the Inner Tracking System (ITS). The large volume TPC serves as the main tracking system and provides particle identification by measuring the particle's energy loss $\rm{d}E/\rm{d}x$ in the detector gas. The ITS consists of two layers each of silicon pixel, silicon strip and silicon drift detectors and provides precise vertex reconstruction and tracking, thus improving the momentum resolution. At forward rapidity, $\jpsi$ mesons are measured through their decay into two muons using the muon spectrometer. It consists of a thick frontal absorber that suppresses most of the hadrons produced during the collision, a tracking system made of 10 planes of Cathode Pad Chambers and an identification system made of 4 planes of Resistive Plate Chambers (RPC) located further downstream behind an iron absorber. In $\pp$ collisions, the RPC are also used to trigger on collisions in which a muon is produced at forward rapidity. Additionally, the V0 detectors, consisting of two scintillator arrays at backward ($-3.7<\eta<-1.7$) and forward ($2.8<\eta<5.1$) rapidities provide the minimum bias trigger (together with the two ITS pixel layers) in both analysis, as well as the centrality determination in $\pbpb$ collisions. All results presented here are inclusive and contain contributions from both direct $\jpsi$ production and feed-down from higher mass quarkonia as well as B hadron decays. 

\begin{figure}[htcb]
\begin{tabular}{cc}
\includegraphics[height=63mm]{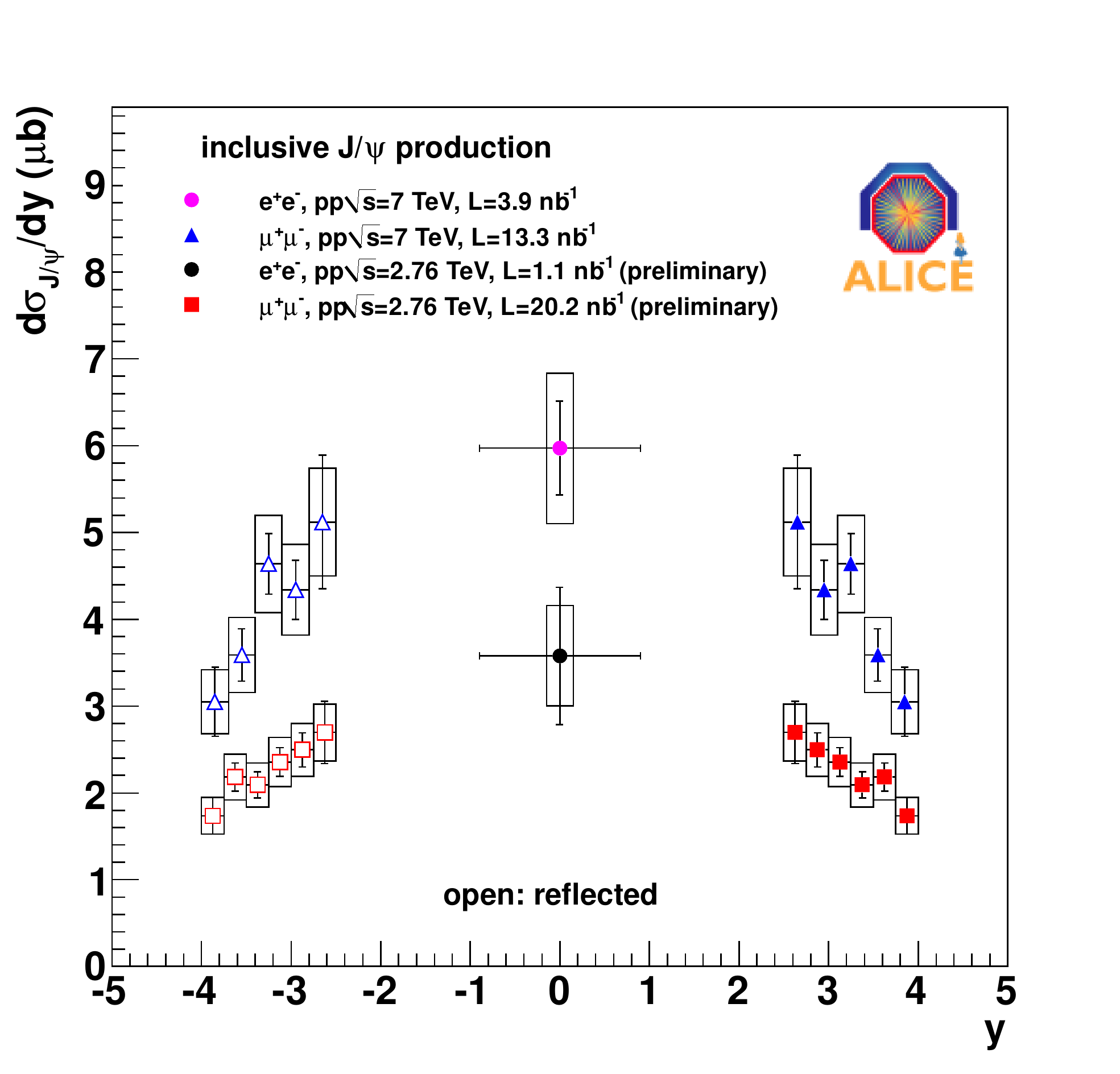}&
\includegraphics[height=63mm]{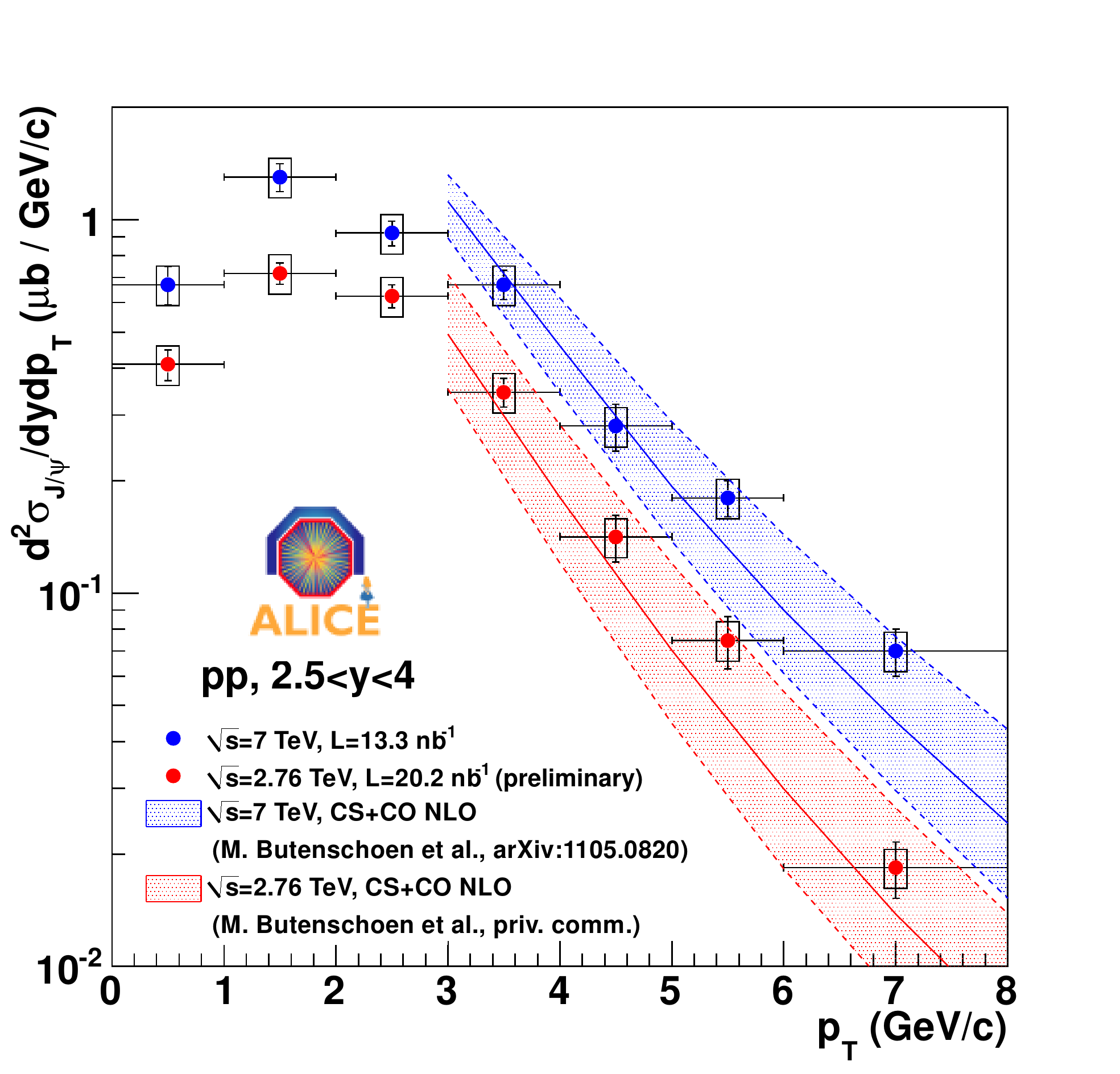}
\end{tabular}
\caption{\label{fig_pp}Differential $\jpsi$ production cross-section as a function of the $\jpsi$ rapidity (left) and transverse momentum (right) in $\sqrt{s}=7$~\cite{ALICE:2011} and $2.76$ TeV $\pp$ collisions. Vertical bars are for statistical uncertainties and boxes are for systematic uncertainties. An additional scaling uncertainty is associated to the total inelastic $\pp$ cross-section and the minimum bias trigger efficiency. It is $4$~\% for the $7$~TeV data, and $7$~\% for the $2.76$~TeV data. Colored lines and bands in the right panel correspond to a NRQCD calculation at next-to-leading order~\cite{Butenschoen:2011}, and its corresponding theoretical uncertainties.}
\end{figure}

The $\jpsi$ differential production cross-section has been measured in $\sqrt{s}=7$~\cite{ALICE:2011} and 2.76 TeV $\pp$ collisions as a function of the $\jpsi$ rapidity (left panel of Figure~\ref{fig_pp}) and transverse momentum (right panel of Figure~\ref{fig_pp}, at forward rapidity). The $\jpsi$ production cross-section as a function of $\pt$ has also been measured at mid-rapidity in 7 TeV $\pp$ collisions. On the right panel of Figure~\ref{fig_pp}, the data are compared to a NRQCD calculation~\cite{Butenschoen:2011} and a reasonable agreement is observed. A good agreement is also observed when comparing these ALICE measurements to the ones performed by the other three LHC experiments: ATLAS, CMS and LHCb.

\begin{figure}[htcb]
\includegraphics[height=60mm]{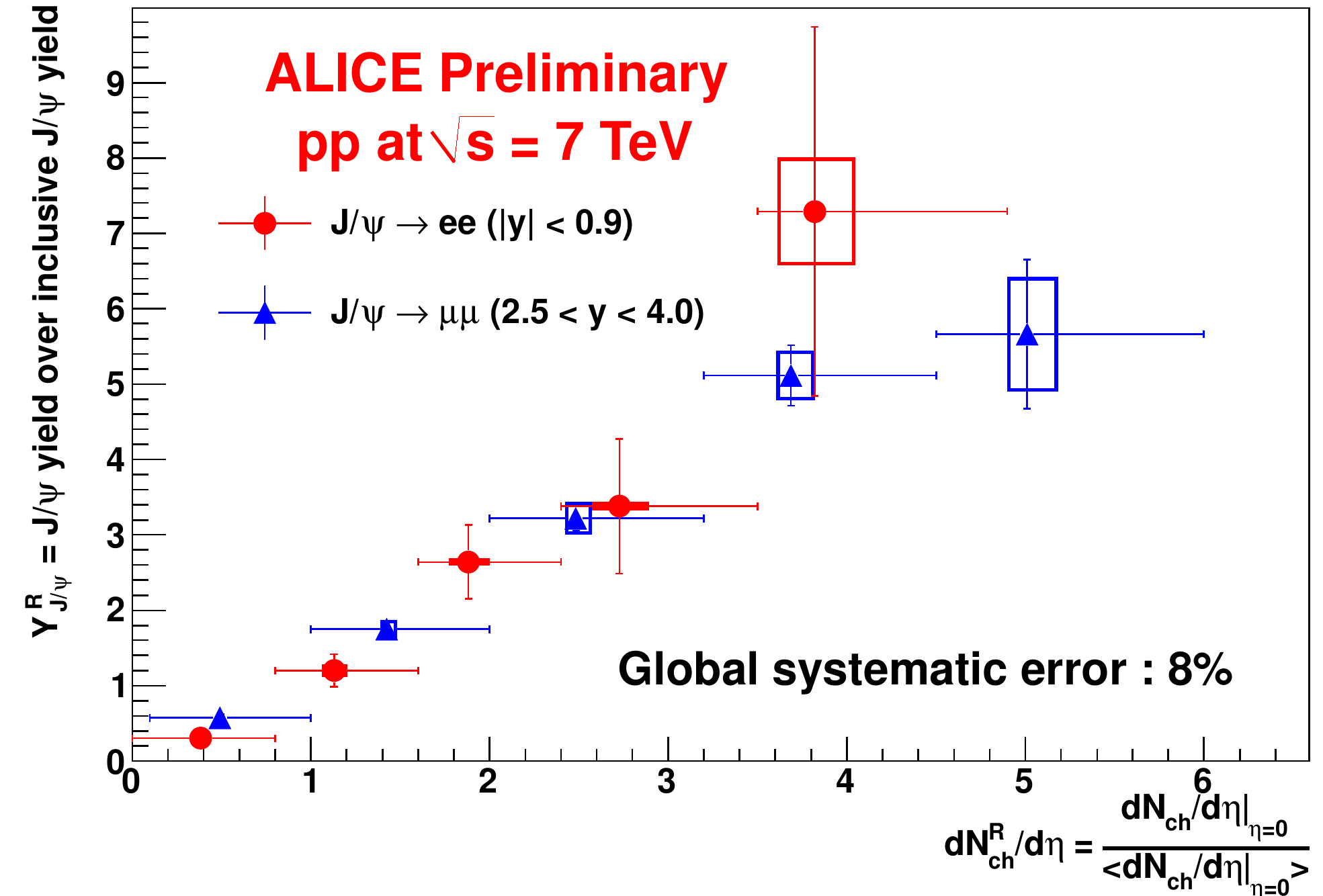}
\caption{\label{fig_mult}Relative $\jpsi$ yield per $\sqrt{s}=7$~TeV $\pp$ collision as a function of the relative charged particles yield at mid-rapidity.}
\end{figure}

The ALICE collaboration has also measured the mean number of produced $\jpsi$ per $\pp$ collision in several bins of the collision's charged particle multiplicity. This measurement is expected to give insight on the dynamics of a $\pp$ collision at LHC energies as well as on the interplay between hard and soft processes in the context, for instance, of multi-parton interactions~\cite{Bernhard:2008}. Results obtained at both mid- and forward-rapidities are presented in Figure~\ref{fig_mult}. The charged particle multiplicity is estimated event by event using the number of track segments reconstructed in the two ITS pixel layers at a pseudo-rapidity $|\eta|<0.6$~\cite{ALICE:2010}. On both axis, the mean number of produced $\jpsi$ and charged particles in each bin are divided by their respective minimum bias values. An approximately linear increase of the $\jpsi$ multiplicity is observed at both rapidities. The interpretation of this observation is still being investigated theoretically.

\begin{figure}[htcb]
\includegraphics[height=60mm]{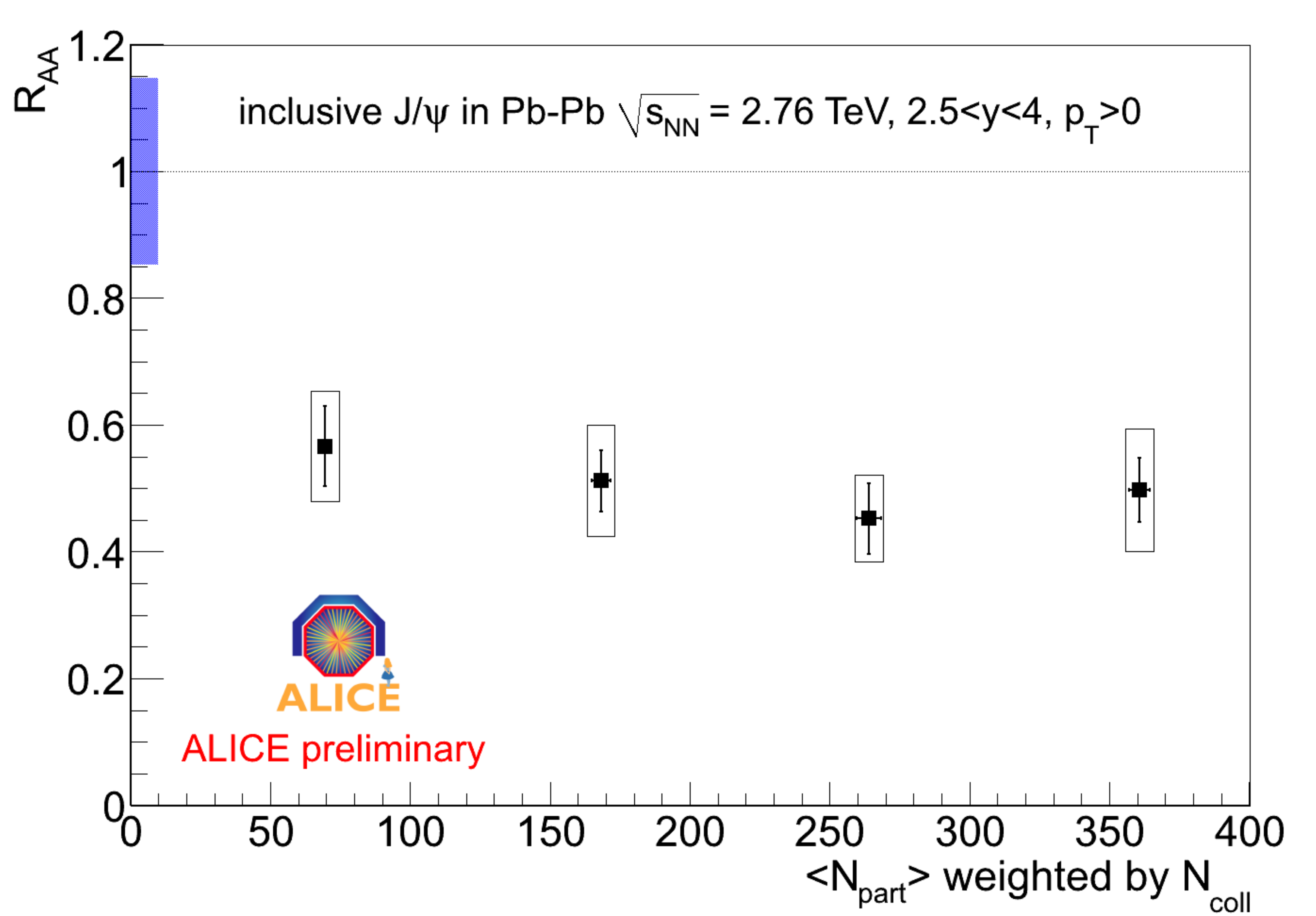}
\caption{\label{fig_raa}The $\jpsi$ nuclear modification factor $\raa$ measured at forward-rapidity in $\snn=2.76$ $\pbpb$ collisions as a function of the number of participing nucleons $\npart$. To correct for biases due to the use of large centrality bins, the mean value of $\npart$ in each bin has been evaluated using the number of binary collisions $\ncol$ as a weight.}
\end{figure}

The first $\pbpb$ collisions at an energy per nucleon-nucleon collision $\snn=2.76$~TeV have been recorded in fall 2010. Comparing the $\jpsi$ data obtained in these conditions to the one measured in $\pp$ collisions at the same energy allows one to form the $\jpsi$ nuclear modification factor $\raa$ for which a value equal to unity is expected in absence of any medium induced modification of the $\jpsi$ production. Figure~\ref{fig_raa} shows the $\jpsi$ $\raa$ measured at forward-rapidity in $\snn=2.76$ TeV $\pbpb$ collisions as a function of the number of nucleons participating to the collision $\npart$, for $\jpsi$ transverse momenta down to zero. A significant suppression is observed already for peripheral collisions (small values of $\npart$), where $\raa$ equals $0.55$. $\raa$ remains essentially constant for increasing collision centralities. These data have been compared to similar measurements performed at RHIC in $\snn=0.2$~TeV $\auau$ collisions~\cite{PHENIX:2011}. For central collisions, the $\jpsi$ suppression is smaller at the LHC than at RHIC. However, due to the different collision energies, the modifications of the parton distribution functions as well as the energy density corresponding to a given $\npart$ are different for the two measurements. Moreover, the final state nuclear absorption of the $\jpsi$ is known to play a significant role at RHIC~\cite{PHENIX:2008} whereas it is expected to be negligible at the LHC. The ALICE data have also been compared to several calculations that include one or several of the following components: modifications of the parton distribution functions, contribution from B meson decays, suppression of the prompt $\jpsi$ via color screening in a QGP, statistical or kinetic formation of $\jpsi$ in a QGP out of uncorrelated charm quarks~\cite{Lansberg:2011,Rapp:2011,Zhuang:2011,Andronic:2011}. Although uncertainties on these calculations are rather large, notably due to the uncertainty on the charm production cross-section, a reasonable agreement is achieved when several of these contributions are accounted for simultaneously. The $\jpsi$ central to peripheral modification factor $\rcp$ has also been measured at both forward- and mid- rapidity, which allows comparisons to other LHC experiments. However, differences in the $\jpsi$ $\pt$ and rapidity range, as well as large statistical uncertainties for the mid-rapidity point prevents to draw firm conclusions from such comparisons.

In summary, the inclusive $\jpsi$ production has been measured both in $\pp$ and $\pbpb$ collisions. In $\pp$, precise measurement of the $\jpsi$ differential cross-section as a function of the $\jpsi$ rapidity and transverse momentum constrain models of the $\jpsi$ production. An approximately linear increase of the number of $\jpsi$ produced per $\pp$ collision is observed as a function of the collision's charged particle multiplicity at mid-rapidity, which has yet to be explained theoretically. In $\pbpb$, the $\jpsi$ nuclear modification factors $\raa$ and $\rcp$ have been measured as a function of the collision centrality. At forward rapidity, a significant suppression of the $\jpsi$ yield with respect to the $\pp$ case is observed, with little dependence on centrality.  
\bibliographystyle{aipproc}
\bibliography{544_PereiraDaCosta}

\end{document}